# ART: Artifact Removal Transformer for Reconstructing Noise-Free Multichannel Electroencephalographic Signals

Chun-Hsiang Chuang*, Kong-Yi Chang, Chih-Sheng Huang, Anne-Mei Bessas

*Abstract*—Artifact removal in electroencephalography (EEG) is a longstanding challenge that significantly impacts neuroscientific analysis and brain–computer interface (BCI) performance. Tackling this problem demands advanced algorithms, extensive noisy-clean training data, and thorough evaluation strategies. This study presents the Artifact Removal Transformer (ART), an innovative EEG denoising model employing transformer architecture to adeptly capture the transient millisecond-scale dynamics characteristic of EEG signals. Our approach offers a holistic, end-to-end denoising solution for diverse artifact types in multichannel EEG data. We enhanced the generation of noisy-clean EEG data pairs using an independent component analysis, thus fortifying the training scenarios critical for effective supervised learning. We performed comprehensive validations using a wide range of open datasets from various BCI applications, employing metrics like mean squared error and signal-to-noise ratio, as well as sophisticated techniques such as source localization and EEG component classification. Our evaluations confirm that ART surpasses other deep-learning-based artifact removal methods, setting a new benchmark in EEG signal processing. This advancement not only boosts the accuracy and reliability of artifact removal but also promises to catalyze further innovations in the field, facilitating the study of brain dynamics in naturalistic environments.

## I. INTRODUCTION

ELECTROENCEPHALOGRAPHY (EEG) is a non-invasive neurophysiological method that records electrical activity in the brain through sensors on the scalp. Known for its millisecond-scale temporal resolution and ability to capture transient brain dynamics, EEG is widely used in neuroscience, cognitive neuroscience, clinical and medical fields, brain–computer interfaces (BCI) [1], sports science, and education studies. Additionally, EEG plays a key role in the development of mobile brain and body imaging [2] techniques, which investigate brain dynamics supporting natural cognition. The portability of EEG hardware enables research beyond traditional laboratories, facilitating the exploration of brain dynamics in natural settings and expanding research scope.

However, EEG signals are highly susceptible to artifacts [3-5] due to their weak amplitudes, typically in the microvolt range. This vulnerability makes them prone to external interferences and significantly affects the signal-to-noise ratio (SNR), challenging data integrity and reliability. Although appropriate shielding and filtering can mitigate such interferences, contamination by physiological artifacts remains a common challenge, especially in naturalistic settings where real-world behaviors are monitored. Artifacts from eye movements or blinks, which manifest as sharp spikes or slow-wave patterns [6], frequently overshadow the EEG signals. Similarly, muscle contractions during physical or stressful activities often introduce high-frequency components [7] that overlap with EEG signals. Improperly positioned electrodes can also lead to problems. For instance, electrodes placed over superficial veins or arteries may inadvertently record heartbeat-related electrical activities, evident as QRS complexes [8] in EEG recordings. Additional problems include poor electrode contact, lead movement, and inherent noise within EEG systems, all of which contribute to channel noise [9]. Given the prevalence of various artifacts and their combinations that overwhelm true neural signals, it is essential to develop advanced countermeasures to simultaneously remove these undesirable artifacts and effectively reconstruct clean data.

In the rapidly evolving fields of EEG analysis and EEG-based BCI technology, the integration of deep-learning algorithms has catalyzed transformative changes [10]. The surge in advanced data analysis capabilities, coupled with advancements in computational resources, has facilitated significant progress in various applications. These include the classification of EEG signals [11, 12], enhancement of processing tools [13, 14], and data generation [15, 16]. One of the core strengths of deep learning in EEG applications is its ability to effectively extract deep feature representations from complex time-series structures [17, 18]. In EEG signal processing, the domain of artifact removal, inspired by developments in computer vision, continues to be dominated

This work was supported by the National Science and Technology Council, Taiwan (NSTC 113-2636-E-007-005 & 113-2221-E-007-121-MY3).

C.-H. Chuang and K.-Y. Chang are affiliated with the Research Center for Education and Mind Sciences and the Institute of Information Systems and Applications at National Tsing Hua University, Hsinchu, Taiwan. Correspondence can be directed to C.-H. Chuang at ch.chuang@ieee.org.

C.-S. Huang is associated with the Product Development Division at ELAN Microelectronics Corporation, Hsinchu, Taiwan; the College of Artificial Intelligence, National Yang Ming Chiao Tung University, Hsinchu, Taiwan; and the College of Electrical Engineering and Computer Science, National Taipei University of Technology, Taipei, Taiwan.

A.-M. Bessasa is with the Research Center for Education and Mind Sciences, National Tsing Hua University, Hsinchu, Taiwan.

by convolutional neural network (CNN) architectures [19, 20]. Numerous studies have used CNNs to map noise-contaminated EEG signals to clean signals in an end-to-end processing manner. While EEG data structurally differ from typical pixel-based images, various efforts have been made to adapt these signals for CNN processing. For example, Sun *et al.* [21] implemented a one-dimensional residual CNN (1D-ResCNN) with three multi-scale residual blocks to capture the nonlinear characteristics of EEG signals, though this method faced overfitting issues, especially in removing electromyogram (EMG) artifacts. In response, Zhang *et al.* [22] introduced a novel CNN architecture featuring multiple convolutional layers paired with a single fully connected layer aimed specifically at improving EMG artifact removal.

Recent studies have also explored the potential of generative adversarial networks (GANs) [23] to further enhance EEG signal processing. This approach has been applied to various tasks, including augmenting data for BCIs, increasing EEG resolution through super-sampling, and repairing damaged EEG data segments [24]. Luo *et al.* [25] redesigned the loss function using the Wasserstein distance, enabling the GAN to effectively reconstruct EEG data across varying sampling rates and sensitivities. This adjustment led to a boost in the accuracy of classification tasks following signal reconstruction. Additionally, Sawangjai *et al.* [26] introduced EEGANet, a GAN-based model designed to restore multichannel EEG signals contaminated by ocular artifacts. This approach achieves SNR results comparable to those of eye-blink detection algorithms, which typically rely on additional electrooculography (EOG) reference channels for enhanced accuracy.

To further enhance the ability of artifact removal techniques to capture the temporal dependencies in EEG signals, techniques originally developed for natural language processing, such as recurrent neural networks (RNNs) and long short-term memory (LSTM) networks [27, 28], have been adopted. These approaches allow for a detailed analysis of dynamic wave patterns in EEG by using sequential processing methods similar to those used in language modeling. This capability is essential, considering the rapid fluctuations in EEG data, which can vary on a millisecond-scale and involve thousands of samples per second. Zhang *et al.* [29] demonstrated the feasibility of using RNNs for end-to-end EEG denoising. McIntosh *et al.* [30] applied RNNs to reduce ballistocardiogram (BCG) artifacts in EEG data collected during simultaneous functional magnetic resonance imaging (fMRI) sessions. Extending these applications, Gao *et al.* [31] introduced DuoCL, a hybrid model combining CNNs with LSTM networks to better capture the morphological features of EEG time-series, illustrating the advanced integration of deep-learning techniques for enhanced EEG signal processing. These studies highlight the versatility of RNNs in handling complex noise types in EEG recordings under diverse experimental conditions.

The burgeoning success of transformer [32] architectures has been extended to EEG signal processing and BCIs, particularly in artifact removal. Many researchers have explored the integration of attention mechanisms and transformer architectures for EEG data. Pu *et al.* [33] proposed EEGDet, a model that combines a 1-D denoising network with a 2-D transformer to enhance feature integration by leveraging local and non-local information through self-attention blocks to capture self-similarities and feed-forward blocks for broader data assimilation. Yin *et al.* [34] introduced GCTNet, a GAN-based architecture, whose generator uses parallel CNN and transformer blocks to effectively capture both local and global temporal dependencies, while discriminator reduces discrepancies between clean and denoised EEG signals for more precise artifact removal.

Despite recent advances, most existing methods treat EEG channels individually, potentially overlooking the intricate interconnections among them. This approach might not fully capture the unique spatial characteristics of each channel, which are crucial for understanding the comprehensive spatial dynamics of EEG data. Given the extensive foundational studies of brain networks [35] and the proven enhancement of BCI performance through spatial-temporal features [36, 37], leveraging a multichannel approach holds significant promise for improving artifact removal techniques. By incorporating the intricate relationships among EEG channels, this approach can exploit the comprehensive spatial dynamics inherent to multichannel EEG data, potentially optimizing the accuracy and efficacy of the artifact removal process. This concept was demonstrated in prior work by Chuang *et al.* [38], who developed IC-U-Net, based on the U-Net architecture [39], to map the relationships between noisy input mixtures and their corresponding clean targets. This method effectively used multichannel EEG time-series to compute losses, showcasing the benefits of leveraging spatial dynamics across multiple channels for artifact removal.

To better harness spatial information from EEG time-series, prior studies [40, 41] have explored the incorporation of attention mechanisms to model inter-channel dynamics. Integrating this approach into artifact removal models could be a key strategy for enhancing the ability to discern relevant patterns across multiple channels. Therefore, this study first introduces a self-attention mechanism into the original IC-U-Net architecture to enhance signal reconstruction. Second, it presents ART, a novel multichannel EEG denoising technique inspired by advanced neural architectures such as Google's Vision Transformer (ViT) [42] and OpenAI's Whisper [43]. ART leverages the transformer framework, known for its ability to model dependencies within data, aiming to improve EEG signal denoising. Additionally, ART adopts a non-autoregressive decoding strategy to increase processing speed and facilitate potential real-time EEG-based BCI applications.

In addressing the challenges of EEG artifact removal, it is crucial for models to effectively map from artifact-contaminated data to clean data. However, perfectly clean EEG data are not publicly available. To overcome this, our study, building on previous work [38], has generated a significant dataset of synthetic noisy–clean EEG pairs using independent component analysis (ICA) [44-46] to mix brain and non-brain components. This strategy enables the training of sophisticated denoising models and enhances the accuracy of EEG signal reconstruction. We expect this strategy to also be suitable for ART.

The contributions of this study can be summarized as follows:

- Novel EEG Denoising Models: This study introduces two innovative models. First, we enhance the IC-U-Net by incorporating self-attention mechanisms into its U-Net architecture, significantly improving its ability to capture interactions among EEG channels during signal reconstruction. Second, the ART model uses the transformer architecture, effectively managing multichannel EEG data. This model excels at capturing the millisecond-scale transient dynamics of EEG, providing a comprehensive, end-to-end solution for denoising various types of artifacts in a single step, with deep-learning layers specifically adapted to EEG characteristics.
- Data Preparation: This study refines the approach to generating noisy–clean EEG data pairs using ICA. By backprojecting selected independent components from source to sensor levels, we construct unique noisy and clean EEG pairs. This process involves the spatial mixing matrix from ICA to blend brain and non-brain components, creating realistic training scenarios. Each mixing matrix, tailored to its respective components, facilitates robust generation of training data, essential for training the ART model under supervised learning conditions. This approach could also be applied to train other artifact removal techniques.
- Comprehensive Validation: To validate the proposed models, extensive testing is conducted across a broad spectrum of open datasets from diverse experiments and BCI applications, such as motor imagery, steady-state visually evoked potentials, and simulated driving tasks. The evaluation involves traditional metrics such as mean squared error (MSE) and SNR, as well as innovative approaches like source localization and EEG component classification to ascertain the brain signal integrity in reconstructed data. Additionally, the performance of BCI systems using these reconstructed signals is assessed, comparing different deep-learning-based artifact removal techniques.

In comprehensive evaluations, ART consistently outperforms other deep-learning-based artifact removal techniques selected for comparison.

## II. METHODS

This study presents a novel deep-learning architecture for EEG signal reconstruction designed to significantly suppress artifacts and enhance the SNR, thereby boosting BCI performance and improving data interpretation. In addition to using a blind source separation (BSS)-based method for generating noisy–clean training data (Figure 1A), we have developed two series of models based on the deep denoising autoencoder (DDAE) framework, which clean noisy EEG signals in an end-to-end manner, leveraging advanced encoder and decoder architectures. The first model series introduced is the IC-U-Net series, as illustrated in Figure 1B. Extending the previously proposed models, such as IC-U-Net [38] and IC-U-Net++ [47], a new variant, IC-U-Net-Attn, integrates the self-attention mechanism into the U-Net framework, enhancing its feature extraction capabilities. The second model series, named ART (Figure 1C), employs a transformer architecture. This method is highly regarded for its effectiveness in sequence-to-sequence feature extraction and is expected to excel at capturing the complex temporal structures inherent in EEG signals.

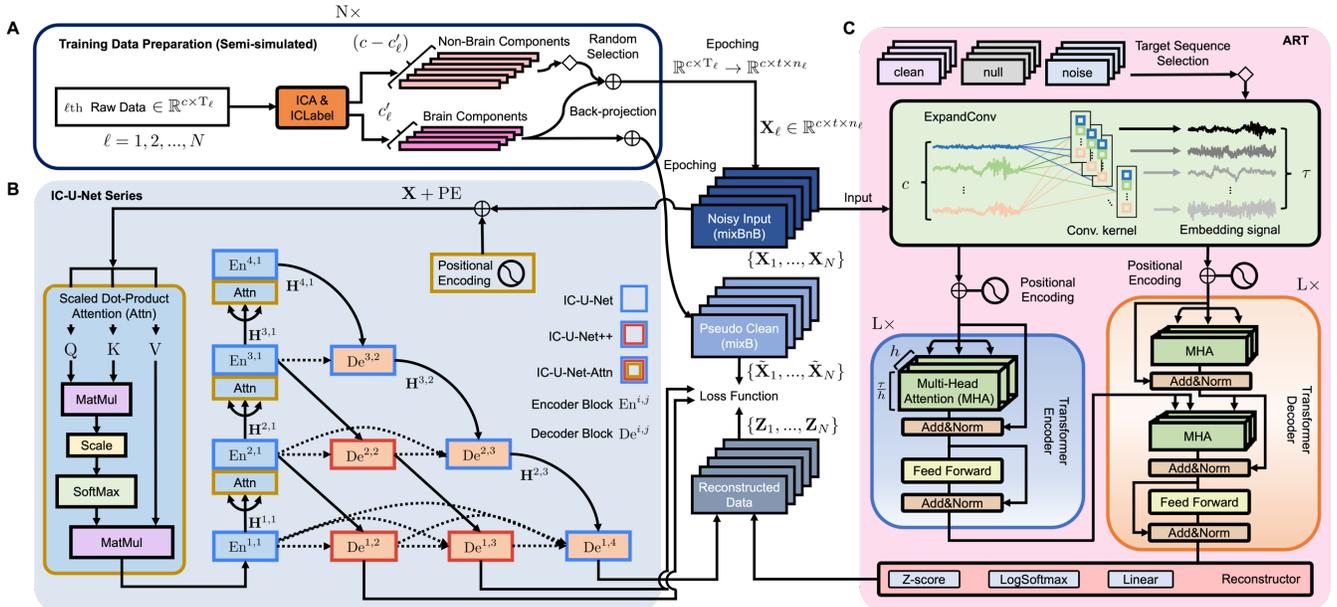

**Fig. 1.** Overview of signal reconstruction methods, IC-U-Net and ART series, designed to counteract artifacts. A. Training data preparation involves BSS and source mixing to create synthetic noisy inputs and pseudo-clean data, using ICA and ICLabel. B. IC-U-Net (blue boxes), and its variants: IC-U-Net++ (blue and red boxes) and IC-U-Net-Attn (blue, red, and orange boxes). C. ART series, categorized into $ART_{clean}$, $ART_{null}$, and $ART_{noise}$, differentiated by their target sequences—pseudo-clean data, null matrix, and noisy input, respectively.

Given $N$ $c$-channel datasets in the EEG databank, each raw dataset has a dimensionality of $\mathbb{R}^{c \times T_\ell}$, where $T_\ell$ represents the number of data points and $\ell$ ranges from 1 to $N$. Each dataset is processed using ICA [44, 45] and ICLabel [46] to decompose the samples into $c_\ell$ brain sources and $c - c_\ell$ non-brain sources, such as ocular and muscular artifacts. Subsequently, the noisy training input is constructed by mixing all $c_\ell$ brain sources with a randomly selected number of non-brain sources, ranging from at least one to at most $c - c_\ell$. Pseudo-clean data are generated by combining all $c_\ell$ brain sources. Both mixtures are achieved through back-projection, using their respective ICA weights. Next, these newly mixed data are epoched into $n_\ell$ segments, each with a uniform length of $t$. This process is applied to each dataset, resulting in the collection of noisy input and pseudo-clean data, denoted as $\{\mathbf{X}_1, \mathbf{X}_2, \ldots, \mathbf{X}_N\}$ and $\{\widetilde{\mathbf{X}}_1, \widetilde{\mathbf{X}}_2, \ldots, \widetilde{\mathbf{X}}_N\}$, respectively. Each dataset maintains the dimensionality of $\mathbb{R}^{c \times t \times n_\ell}$. Additionally, the total number of segments across all datasets is $\sum_{\ell=1}^{N} n_\ell = M$.

The proposed models, IC-U-Net or ART, are engineered to establish a mapping between noisy inputs and pseudo-clean data, thereby generating the reconstructed data, denoted $\mathbf{Z}$. These models are trained with loss functions designed to minimize the disparity between the reconstructed data and corresponding pseudo-clean data.

*A. Network Architecture of IC-U-Net Series*

*IC-U-Net*

The IC-U-Net series (Figure 1B) consists of three models that progressively increase in complexity by integrating additional functionalities. These enhancements range from adding multiple layers with shortcuts that connect different parts of the network to incorporating an attention mechanism to dynamically emphasize the most critical features needed for signal reconstruction. This progression is visually represented by different colored boxes in Figure 1B.

The first introduced model is the IC-U-Net [38], which features encoder blocks ($\mathrm{En}^{i,j}$) and decoder blocks ($\mathrm{De}^{i,j}$). These blocks are visually distinguished in Figure 1B: Encoder and decoder blocks are marked with blue boxes against light blue and orange backgrounds, respectively. The indices $i$ and $j$ represent the depth and width dimensions, respectively, within the network architecture. Given that the number of EEG datapoints often follows a power of two, the outputs of $\mathrm{En}^{i,j}$ and $\mathrm{De}^{i,j}$ in the model are represented as $\mathbf{H}^{i,j} \in \mathbb{R}^{(\lambda \times 2^i) \times (t/2^i)}$, where $\lambda$ denotes the dimension of the initial feature map. In this study, the dimensionality of the input data is adjusted through an initial convolution: The channel dimension ranges from $c$ to $\lambda$, with $\lambda$ set as 32.

We consider the example of a four-layer structure containing four encoder blocks and three decoder blocks. Each encoder block $\{\mathrm{En}^{i,1} | i = 1,2,3,4\}$ performs double convolution to generate a set of feature maps followed by batch normalization and rectified linear unit (ReLU) activation. Subsequently, 1D-maxpooling with a non-overlapping window and stride of 2 is applied, and the output is downsampled by a factor of 2, resulting in $\{\mathbf{H}^{i,1} | i = 1,2,3,4\}$. The decoder blocks can be represented as $\{\mathrm{De}^{i,j} | (i,j) = (1,4), (2,3), (3,2)\}$. These blocks first upsample the encoded feature maps from the previous layer and concatenate them with the encoder features of the same level. These upsampled feature maps undergo convolution with the corresponding decoder to produce dense feature maps, followed by batch normalization. The outputs of the decoder blocks are feature maps denoted as $\mathbf{H}^{3,2}$ and $\mathbf{H}^{2,3}$, with the final output being $\mathbf{Z}_i$, and the reconstructed data corresponding to the $\ell^{\mathrm{th}}$ raw dataset.

*IC-U-Net++*

Expanding on IC-U-Net and building upon U-Net++[48], the second model, IC-U-Net++, incorporates dense skip connections and introduces additional decoder blocks between existing layers (red boxes in Figure 1B) to enhance the model's capability of resolving the EEG structure. This addition helps optimize signal handling through multiple layers of shortcuts that connect different parts of the network. These shortcuts enhance the network's ability to learn and retain details by allowing information to flow freely, thereby facilitating a more detailed and refined reconstruction of data. Notably, this enhanced model generates three outputs through $\mathrm{De}^{1,2}$, $\mathrm{De}^{1,3}$, and $\mathrm{De}^{1,4}$, which are used to evaluate their discrepancies with the pseudo-clean data. The model is trained using the average loss calculated across its three outputs. Finally, the output derived from $\mathrm{De}^{1,4}$ is selected as the reconstructed data $\mathbf{Z}$.

*IC-U-Net-Attn*

While IC-U-Net and IC-U-Net++ are capable of end-to-end multichannel signal reconstruction, they rely on one-dimensional convolution that operates on individual channels. This approach focuses on intra-channel feature extraction and may neglect important inter-channel relationships, which are crucial for understanding complex signal dynamics. To address this limitation, this study expands on IC-U-Net++ by developing the IC-U-Net-Attn model, which integrates positional encoding and a self-attention mechanism ($\mathrm{Attn}$). This enhancement aims to capture and leverage inter-channel relationships that are overlooked by the simpler one-dimensional convolutions in previous models.

Specifically, before the first encoder block, $\mathrm{En}^{1,1}$, the model processes noisy data $\mathbf{X}$ by adding sine and cosine functions, $\mathrm{PE}(2q, p) = \sin(p/10000^{2q/c})$ and $\mathrm{PE}(2q+1, p) = \cos(p/10000^{2q/c})$, respectively, where $p$ indexes the position in the time-series, ranging from 0 to $t-1$, and $q$ maps the dimension index, spanning from 0 to $c/2 - 1$. The data are then resized from $c$ to $\lambda$, represented as $(\mathbf{X} + \mathrm{PE})'$, and further processed by the $\mathrm{Attn}$ module. This module incorporates QKV-attention, which includes matrix multiplication, scaling, and softmax operations on Q, K, and V to capture the spatial dependencies among channels. This process is expressed as $\mathrm{softmax}\left(\frac{\mathrm{QK^T}}{\sqrt{c}}\right)\mathrm{V}$, where $\mathrm{Q} = \mathrm{K} = \mathrm{V} = (\mathbf{X} + \mathrm{PE})'$. Similar PE and $\mathrm{Attn}$ processes are applied prior to each subsequent encoder block, $\{\mathrm{En}^{i,1}|i=2,3,4\}$, with $q$ spanning from 0 to $(\lambda \times 2^i)/2 - 1$. This upper limit varies according to the dimension of each processed feature map.

Incorporating positional encoding and self-attention mechanisms into the artifact removal model enhances the learning of complex interdependencies within and across EEG signal channels, leveraging their inherent characteristics.

*B. Network Architecture of IC-U-Net Series*

Figure 1C illustrates the overview of the ART model structure. This transformer-based model includes blocks for expanding convolution, positional encoding, encoding, target sequence selection, decoding, and reconstruction.

*Expanding convolution and transformer*

Each EEG input, featuring $c$ channels and $t$ data points, undergoes the ExpandConv operation to enhance the embedding dimension from $c$ to $\tau$. This is accomplished using convolution operations with a kernel size of $1 \times 1$, which expands the dimensionality of the spatial structure and augments the feature representation of the EEG data. In this study, $\tau$ is set as 128. The expanded EEG data are then processed using positional encoding to integrate positional information. This operation allows the model to understand the relative positions of the EEG time-series, facilitating learning of patterns and trends that may be dependent on the temporal or spatial context. The operation uses parameter $\tau$ instead of $c$, altering the dimension index $q$ to range from 0 to $\frac{\tau}{2} - 1$. The expanded EEG data, now integrated with sine and cosine functions, is subsequently fed into the Encoder and Decoder blocks for further processing.

The encoder consists of $L$ identical blocks, each equipped with a multi-head attention (MHA) layer designed to process EEG data in parallel. Each MHA layer uses $h$ heads, each focusing on distinct EEG signal segments sized $\frac{\tau}{h} \times t$. This allows for a comprehensive analysis of EEG segments from multiple perspectives. Each head generates query, key, and value vectors through linear projections, calculates scaled dot-product attention, and then concatenates the outputs from all heads to restore the original dimension $\tau \times t$. This process concludes with another linear projection to complete the MHA operation. The number of heads is set as $h = 8$ in this work.

Residual addition surrounds the sublayers within each stack of the encoder, followed by layer normalization, denoted as Add&Norm in Figure 1C, to maintain signal integrity and processing stability. The feed-forward layer completes the encoding process. Input features undergo a linear projection, expanding from $\tau$ to $\tau'$, followed by activation through a ReLU function. Subsequently, another linear projection compresses the features back to the original dimension $\tau$, completing the process. The output of the encoder then serves as the key and value vectors for the MHA layer within the decoder.

Compared with the encoder, each decoder block features an additional MHA layer designed to process the output data. This layer generates the necessary query vector, which is then fused with the encoder's output through QKV operations and a feed-forward layer. After sequential processing by $L$ identical blocks, the output from the decoder is forwarded to the final layer, the reconstructor, preparing it for the ultimate phase of signal reconstruction.

Throughout the process, the temporal dimensionality, $t$, remains unchanged.

*Decoding and signal reconstruction*

This study implements a non-autoregressive decoding approach to reconstruct signals, focusing on capturing the complete fluctuating structure of EEG time-series. This technique ensures a more accurate depiction of underlying brain activity by effectively managing signal variations. Additionally, the research explores three different target sequence configurations in the ART model to evaluate their impact on performance, providing insights into the most effective methods for artifact-free EEG reconstruction.

The three target sequences for the encoder include pseudo-clean data ($\widetilde{\mathbf{X}}$), zero matrix ($\mathbf{O}$), and noisy input ($\mathbf{X}$). These choices represent different scenarios in terms of data information. Pseudo-clean data correspond to the encoder being trained with data leakage, zero matrix reflects training with no prior information, and noisy input simulates a practical scenario for artifact removal where the input and target sequences are the same non-processed time-series. These variations provide valuable insights into the model performance under different conditions and applications.

In the ART model, two significant modifications are implemented in the encoder and the output block. First, the masking mechanism used in the original transformer's decoder is disabled, liberating the model from accessing future information. This shift from autoregressive to non-autoregressive methodology enables the model to capture temporal dependencies across the entire EEG data segment and generate the complete time sequence in a single step, rather than sequentially producing each temporal point.

In the final block, a critical component termed the reconstructor is introduced. This component involves a series of operations including linear projection, LogSoftmax activation, and z-score normalization. The linear projection layer reduces the signal dimension from $\tau$ back to the original number of channels, $c$. Following this dimension reduction, LogSoftmax activation is applied to enhance numerical stability by moderating extreme values and suppressing burst and erratic patterns. Finally, z-score normalization is applied to scale the data, ensuring that they oscillate around zero,

TABLE 1
DATA USED IN THIS WORK

| Dataset | Experiment | # of EEG Sessions | Device | # of Channels | Sample Rate (Hz) | Data Length (s) | Performance Metric | | | | |
|---|---|---|---|---|---|---|---|---|---|---|---|
| | | | | | | | MSE | SNR | ICLabel (# of non-brain) | Source localization | BCI |
| Chuang et al. [38] | Eyes-Closed/Eyes-Open Resting-State | 546 | NuAmp, Compumedics Neuroscan | 30 | 1000 | 289.3±32.9 | ✓ | ✗ | ✓ | ✗ | ✗ |
| Zhang et al. [29] | EEGdenoiseNet (Semi-synthetic data) | 4514 segments | – | 1 | 256 | 2.0/epoch | ✓ | ✗ | ✗ | ✗ | ✗ |
| Cao et al. [51] | Simulated Driving | 76 | SynAmps2, Compumedics Neuroscan | 32 | 500 | 4737.3±1266.9 | ✗ | ✓ | ✓ | ✗ | ✗ |
| Schalk et al. [49] | Motor Imagery (BCI2000) | 109 | – | 64 | 160 | 996.7±31.2 | ✗ | ✗ | ✓ | ✓ | ✓ |
| Liu et al. [50] | Visual Speller (SSVEP) | 70 | SynAmps2, Compumedics Neuroscan | 64 | 250 | 1211.4±132.2 | ✗ | ✓ | ✓ | ✓ | ✗ |
| Gu et al. [54] | Simultaneous EEG-fMRI | 33 | BrainAmp MR series, Brain Products | 32 | 5000 | 6639.6±1828.2 | ✗ | ✗ | ✓ | ✗ | ✗ |
| Miltiadous et al. [60] | Eyes-Closed Resting-State | 88 | Neurofax EEG-2100, Nihon Kohden | 19 | 500 | 802.1±141.1 | ✗ | ✗ | ✓ | ✗ | ✗ |
| Chavarriaga and Millan [55] | Error Processing | 6 | ActiveTwo, Biosemi | 64 | 512 | 1824.3±29.1 | ✗ | ✗ | ✓ | ✗ | ✗ |
| Broderick et al. [56] | Speech-in-Noise | 21 | ActiveTwo, Biosemi | 128 | 512 | 675.0±75.4 | ✗ | ✗ | ✓ | ✗ | ✗ |
| Blankertz et al. [66] | Motor Imagery (BCI Competition IV) | 7 | BrainAmp MR plus, Brain Products | 59 | 1000 | 1905.4±1.0 | ✗ | ✗ | ✓ | ✗ | ✗ |
| Delorme et al. [59] | Go/No-Go | 28 | – | 31 | 1000 | 2597.0±734.1 | ✗ | ✗ | ✓ | ✗ | ✗ |
| Li et al. [52] | Oddball | 13 | ActiveTwo, Biosemi | 79 | 256 | 875.1±110.8 | ✗ | ✗ | ✓ | ✗ | ✗ |

mirroring the natural fluctuations in voltage observed in EEG signals.

*C. Model Training and Training Datasets*

The aforementioned models were trained on the same dataset as in [38] (Table 1), consisting of $N = 546$ EEG sessions. Each session pertains to recordings from 30 channels at a sampling rate of 1,000 Hz, spanning approximately 5 min. The noisy–clean EEG pairs were prepared using ICA [44, 45] and ICLabel [46] (Figure 1A). In preparing clean training data, only those components identified as brain with a confidence level exceeding 80% were deemed clean. These components were then backprojected to the channel domain, resulting in a mixture of brain components (mixB, $\widetilde{\mathbf{X}}$) with the same channel size as the original data, without interference from any non-brain components. Conversely, the noisy training data were generated by backprojecting the previously identified brain components along with one of five predetermined non-brain components, forming a mixture of brain and non-brain components (mixBnB, $\mathbf{X}$). The categories of non-brain components included eye, muscle, heart, channel noise, and other. These mixed time-series were segmented into non-overlapping 4-s subsegments and subjected to z-score normalization, resulting in a total of 132,675 noisy–clean EEG subsegment pairs. This dataset was subsequently divided into training, validation, and test sets, with allocations of 80%, 10%, and 10%, respectively.

The hyperparameters for model training were set as follows. The number of epochs was 60, and the initial learning rate was 0.01. The batch size was set as 128 for the IC-U-Net models and 32 for the ART models. The IC-U-Net used the stochastic gradient descent optimizer, whereas ART employed the Adam optimizer. The MSE between the clean and reconstructed data, i.e., loss, was calculated for training. The entire reconstruction process was conducted on the Pytorch 2.2.1 platform using an Nvidia GeForce RTX 3090 Ti GPU.

*D. Model Testing and Performance Comparison*

The performance of IC-U-Net and ART frameworks in artifact removal was evaluated against several cutting-edge deep-learning models, including 1D-ResCNN [31], DuoCL [31], and GCTNet [34]. These three models were trained using ~90% (i.e., 4,062 EEG samples) of EEGdenoiseNet [29]. The comparison involved a variety of datasets and performance metrics, as elaborated in the subsequent sections. The remaining 10% of the EEGdenoiseNet data (452 samples) was used as the test dataset.

*Datasets*

This study expanded its evaluation of artifact removal models by incorporating 11 publicly available EEG datasets from various classic experiments, as outlined in Table 1. These datasets included recordings from motor imagery [49], steady-state visual evoked potentials (SSVEPs) [50], Go/No-Go tasks [51], oddball sequences [52], and simulated driving and semi-synthetic data [29], collected using diverse EEG devices like Brain Products, Biosemi, and Compumedics Neuroscan. These datasets, known for containing common EEG artifacts such as eye movements and muscle activity, were sourced from different EEG devices to ensure the broad applicability of the findings. This diversity in data sources enabled an assessment of the models' generalizability. Electrode mapping was applied to select electrodes positioned similarly to those specified in the training data configuration [38]. The number of EEG channels to be analyzed was reduced to 30 or fewer.

*Performance metrics*

This study used a range of performance metrics (Table 1) to evaluate various aspects of model performance. The first metric, MSE, measured discrepancies between reconstructed signals and pseudo-clean data from the training data, specifically the resting-state EEG [38]. This metric was pivotal for verifying the models' training convergence and assessing their effectiveness in artifact suppression across various spatial channels, with the lowest possible MSE indicating optimal model performance.

The second performance metric was the SNR, which assesses the quality of a signal relative to the level of background noise. SNR was used in two experiments. The first was a simulated driving task [51], where SNR measured the power ratio between the P300 component signal ("S") and baseline noise ("N"). In the second experiment, involving SSVEPs [50] with 40 flicker frequencies, "S" was defined as the average power of the first three harmonics [53] of the flicker frequency, while "N" represented the average power of the remaining frequencies. The SNR values were converted to decibels (dB) to determine whether the stimulus-evoked waveform was altered or improved post artifact removal.

The third performance metric [38] evaluated the efficacy of artifact removal techniques by analyzing the number of non-brain components in the reconstructed signals. This was achieved through blind source separation techniques, such as ICA combined with ICLabel [46]. ICA segregated the reconstructed signals into distinct components, which ICLabel then classified into brain and non-brain categories. The change in the number of non-brain components provided a measure of the models' effectiveness in artifact suppression. This evaluation was performed across all EEG datasets [38, 49-52, 54-60] enumerated in Table 1, ensuring a thorough assessment of the artifact removal capabilities.

An additional validation method in this study was the precise localization of EEG sources after artifact removal processing. The fourth validation method used source localization techniques to verify alignment with established neuroscience knowledge. This method leveraged the ROIconnect toolbox [61] to acquire voxel time-series, employing linear constrained minimum variance beamforming for precise source projection. Subsequently, region-wise powers were calculated by aggregating voxel powers within specific brain regions. This study selected datasets from motor imagery [49] and SSVEP [50] experiments to examine the presence of well-known EEG responses, namely, the event-related desynchronization of the $\mu$ rhythm (8–12 Hz) in the contralateral motor cortex and stimulus-synchronized activity in the occipital cortex, in the processed signals.

The fifth validation assessed whether denoising models could enhance BCI performance using the motor imagery dataset [49]. The common spatial patterns (CSP) technique was used with a support vector machine (SVM) [62] to build a left/right-hand motor imagery classification system. This system used a 4:1 training-to-testing ratio through the holdout method, repeated 10 times to determine the average performance. The classification accuracy of signals processed by various artifact removal techniques was then compared.

## II. RESULTS

### A. Evaluation of Proposed Artifact Removal Models

*Ablation study on loss convergence*

The first evaluation was to investigate the loss convergence over 60 training epochs in the training, validation, and test phases. The loss metric was the MSE between the reconstructed signal and pseudo-clean data. As shown in Figure 2A, as the number of epochs increases, a universal decrease in MSE is observed across all phases, indicating successful convergence of model training across all proposed methods.

During training, the ART series consistently achieves lower MSE than the IC-U-Net series, indicating more efficient learning. However, the ART models show divergent convergence patterns. $ART_{clean}$ exhibits a rapid decrease in MSE to 0.024 early in training, while $ART_{null}$ starts with a high MSE of 1.891 that reduces significantly, by nearly 50 times, after a few training epochs, matching the performance of the other ART models. $ART_{noise}$ exhibits a steady and intermediate decline in MSE compared with $ART_{clean}$ and $ART_{null}$. Conversely, the IC-U-Net series maintains MSE values approximately 3–9 times higher than those of the ART models, with IC-U-Net-Attn achieving the best MSE of 0.133 within its series. This number is approximately half that of the MSE observed in the original IC-U-Net.

Similar loss convergence patterns are observed during the validation and testing phases. Generally, the IC-U-Net series exhibits somewhat erratic and slightly higher MSE values compared with the training phase, but they remain within a comparable range. $ART_{clean}$'s loss remains stable with an MSE of approximately 0.025. Interestingly, $ART_{noise}$ exhibits the best performance, with its MSE during the validation and test phases surpassing those from the training phase. Overall, the proposed artifact removal techniques, particularly ART, demonstrate efficacy in restoring pseudo-clean signals from noisy data. Detailed MSE values for each phase and model configuration can be found in the supplementary material (Table S1).

*MSEs across spatial locations*

Figure 2B provides the results of channel-based MSE analysis to explore whether the model's learning is uniformly distributed across all channels, with values collected from the 60th training epoch of the test phase.

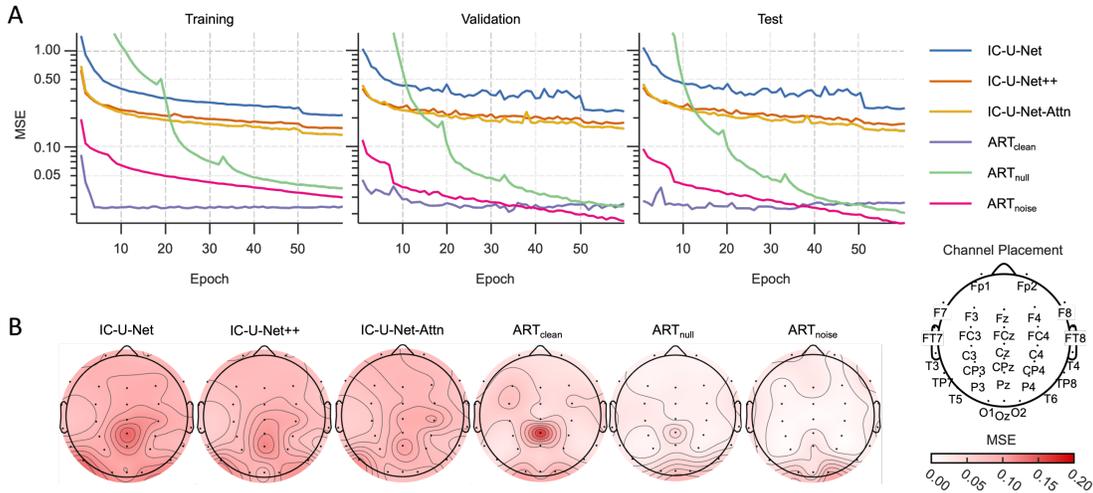

**Fig. 2.** Evaluation of models' efficacy in restoring pseudo-clean EEG signals. A. Loss convergence during the training, validation, and test phases, using MSE to measure discrepancies between reconstructed and pseudo-clean signals. B. Analysis of MSE across EEG channels in the test phase, assessing each model's ability to simultaneously capture spatial and temporal information in an end-to-end manner.

Overall, IC-U-Net++, IC-U-Net-Attn, $ART_{null}$, and $ART_{noise}$ effectively restore EEG signals from noisy inputs to pseudo-clean signals across all channels, revealing uniform performance regardless of spatial location. The averaged MSEs are 0.10 for both IC-U-Net models, while they are significantly lower for the two ART series models (approximately 0.04). Notably, although the MSEs of the original IC-U-Net and $ART_{clean}$ are low, both models exhibit deteriorated performance at the CPz channel (MSE = 0.216±0.628).

*Noise suppression capability*

The analysis presented in Figure 2 demonstrates the effectiveness of the proposed models in restoring clean EEG signals from those contaminated by artifacts. Expanding on this, Figure 3 shows how the reconstructed EEG signals, once cleared of eye or muscle artifacts, closely resemble the pseudo-clean signals. This EEG test dataset contains 3,337 segments contaminated by eye activities and 2,534 segments affected by muscle activities (Table 2).

Figure 3A displays a 4-s EEG segment from the FP2 channel, heavily contaminated by eye activity and its reconstructions by the IC-U-Net and ART series. The original noisy input and pseudo-clean data are shown in gray and black traces, respectively, with reconstructions highlighted in color. All models demonstrate effective noise suppression, efficiently removing low-frequency components ($\leq 3$ Hz) associated with eye artifacts as well as high-frequency components ($\geq 20$ Hz). The reconstructed signals closely resemble the pseudo-clean data, illustrating the models' ability to recover EEG signals across different frequencies. The right panel of Figure 3A presents the MSEs between the reconstructed signals and pseudo-clean signals, with the reconstructions derived from EEG segments contaminated by eye activities. The results indicate that $ART_{noise}$ (MSE = 0.033±0.056) significantly outperforms IC-U-Net (MSE = 0.126±0.122), IC-U-Net++ (MSE = 0.113±0.118), IC-U-Net-Attn (MSE = 0.107±0.116), $ART_{clean}$ (MSE = 0.067±0.037), and $ART_{null}$ (MSE = 0.052±0.068) in handling eye artifacts ($ps < 0.001$).

TABLE 2
MODEL PERFORMANCE IN RESTORING PSEUDO-CLEAN SIGNALS

| | | | | | MSE | | | | | |
|---|---|---|---|---|---|---|---|---|---|---|
| | Mixtures of Components | Training (80%) | Validation (10%) | Test (10%) | IC-U-Net | IC-U-Net ++ | IC-U-Net-Attn | $ART_{clean}$ | $ART_{null}$ | $ART_{noise}$ |
| Pseudo Clean | Brain | 39,605 | 3,592 | 3,693 | – | – | – | – | – | – |
| Noisy Input | Eye | 33,369 | 2,982 | 3,337 | 0.126±0.122 | 0.113±0.118 | 0.107±0.116 | 0.067±0.037 | 0.052±0.068 | 0.033±0.056 |
| | Muscle | 27,429 | 2,416 | 2,534 | 0.091±0.104 | 0.078±0.117 | 0.069±0.107 | 0.065±0.033 | 0.042±0.055 | 0.030±0.053 |
| | Brain + Heart | 1,498 | 75 | 151 | 0.085±0.028 | 0.053±0.016 | 0.042±0.011 | 0.058±0.033 | 0.038±0.008 | 0.025±0.006 |
| | Channel Noise | 12,686 | 1,380 | 1,415 | 0.067±0.049 | 0.056±0.060 | 0.050±0.052 | 0.063±0.039 | 0.034±0.012 | 0.023±0.009 |
| | Other | 36,234 | 3,515 | 3,636 | 0.212±0.216 | 0.200±0.220 | 0.204±0.229 | 0.069±0.037 | 0.088±0.138 | 0.061±0.134 |
| | | | | Average | 0.116±0.058 | 0.100±0.061 | 0.094±0.066 | 0.065±0.004 | 0.051±0.022 | 0.034±0.016 |

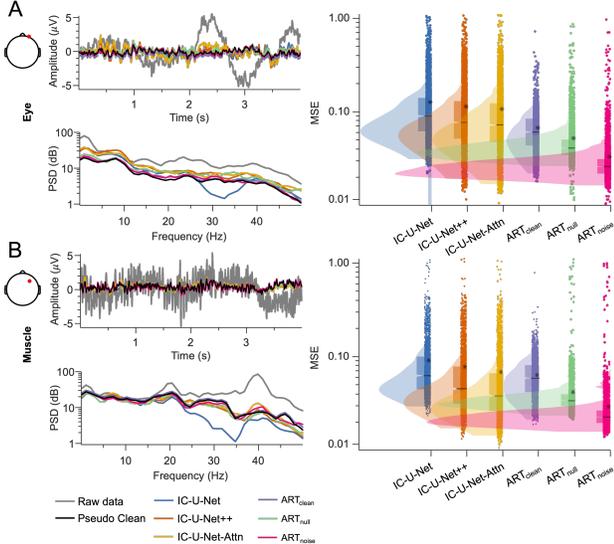

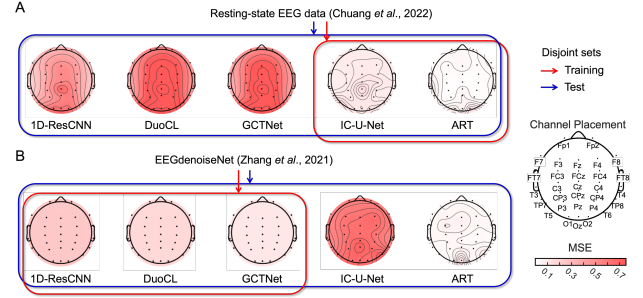

**Fig. 3.** Efficacy of IC-U-Net and ART models in mitigating (A) eye and (B) muscle artifacts. The upper and lower left panels display two 4-s EEG segments of Fp2 and F4 channels affected by artifacts, alongside reconstructions in both time and frequency domains, respectively. The right panel shows a comparative analysis of MSEs, illustrating the model performance in reconstructing signals from noisy inputs. The MSEs from 100 and 200 samples are visualized in a violin plot accompanied by a box plot, where asterisks denote the average, and vertical lines mark the median of MSEs.

Figure 3B illustrates the models' effectiveness in mitigating the impact of high-frequency noise components. The 4-s EEG segment (displayed in gray) contaminated by muscle artifacts is reconstructed, closely matching the pseudo-clean signal in both time and frequency domains. The accompanying MSE values demonstrate $ART_{noise}$'s superior performance (MSE = 0.030±0.053) compared with IC-U-Net (MSE = 0.091±0.104), IC-U-Net++ (MSE = 0.078±0.117), IC-U-Net-Attn (MSE = 0.069±0.107), $ART_{clean}$ (MSE = 0.065±0.033), and $ART_{null}$ (MSE = 0.042±0.055) ($ps<0.001$). Detailed MSE values for reconstructing pseudo-clean data from noisy inputs contaminated with muscle, heart, channel noise, and other artifacts are provided in Table 2. $ART_{noise}$ exhibits superior performance compared with other models in artifact removal ($ps < 0.001$).

Notably, in both cases, the IC-U-Net struggles with signal reconstruction within the frequency range of 25–35 Hz.

*B. Comparison with Other Deep-Learning-based Methods*

*Cross-validation*

Figure 4 presents a comparative analysis of the MSEs for five deep-learning-based techniques, including 1D-ResCNN [21], DuoCL [31], GCTNet [34], IC-U-Net, and $ART_{noise}$. This comparison evaluates the ability of various models to restore clean signals from synthetic noisy input, leveraging datasets used to train both the proposed models and competing methods. To ensure a fair assessment, each model's effectiveness in recovering pseudo-clean or pure signals is verified using the test sets from these datasets. Notably, both the training and testing datasets used for these EEG data sources are completely disjoint.

**Fig. 4.** Cross-validation of five deep-learning-based artifact removal techniques. (A) MSEs for 1D-ResCNN, DuoCL, GCTNet, IC-U-Net, and ART are derived from the resting-state data. (B) MSEs for the same models on the EEGdenoiseNet dataset. All MSE values are derived from test data disjoint from training data. For clarity in subfigure B, MSEs for 1D-ResCNN, DuoCL, and GCTNet are shown as single averaged values across channels.

Figure 4A displays the average MSE results for five different artifact removal methods tested on the proposed models' datasets containing 30 EEG channels. The results indicate that the 1D-ResCNN, DuoCL, and GCTNet achieve MSE ranges of approximately [0.182, 0.704], [0.344, 1.204], and [0.319, 1.097], respectively. Conversely, the IC-U-Net and ART models demonstrate significantly lower MSE, achieving ranges of [0.077, 0.148] and [0.015, 0.071], respectively.

Figure 4B presents the average MSE results using a semi-simulated single EEG channel test dataset (~10%, i.e., 452 EEG samples) from EEGdenoiseNet [29], used to train 1D-ResCNN, DuoCL, and GCTNet. These methods, developed for single-channel artifact removal, yield MSE values represented on a topographic map for clarity, despite using a dataset devoid of spatial information. Conversely, IC-U-Net and ART, which are multichannel artifact removal techniques, map single-channel noisy EEG data to multichannel reconstructed signals. This approach results in varied MSE values across channels, even though the inputs are identical. Compared with the performance depicted in Figure 4A, the results from Figure 4B show that 1D-ResCNN, DuoCL, and GCTNet exhibit enhanced performance on the EEGdenoiseNet dataset, achieving reduced MSEs of 0.256, 0.183, and 0.115, respectively. Conversely, the IC-U-Net's MSE values rise to [0.550, 0.750]. The ART model, demonstrates its robustness in signal restoration across different testing datasets, yielding the lowest MSEs ranging from [0.020, 0.129], thus underscoring its superior efficacy in artifact removal.

*Motor imagery task*

This study further examines how these artifact removal techniques contribute to EEG-based BCI applications. The first selected BCI paradigm is motor imagery, with the dataset corresponding to the well-known BCI2000 [49]. This dataset contains 109 EEG recording sessions, each with 64 channels and sampled at 160 Hz (Table 1). To match the number of channels and sampling rate required in the proposed model, the data are reduced to 30 channels that best match the channel locations of the ART's training data. The data are then upsampled to 256 Hz and processed using a 1–50 Hz band-pass filter. The numbers of EEG samples are 2,480 and 2,438 for left- and right-hand motor imagery, respectively.

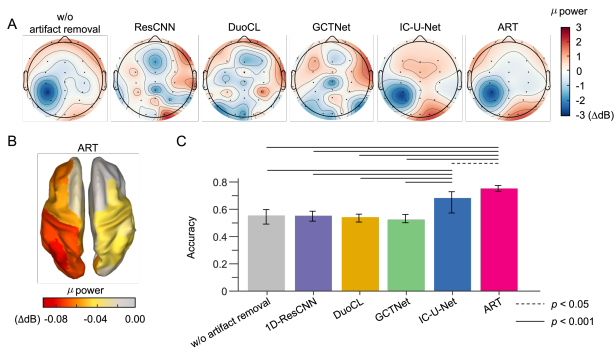

**Fig. 5.** Efficacy of different artifact removal techniques in terms of BCI performance for left-hand and right-hand motor imagery tasks. (A) Topographic maps illustrating $\mu$ band power distributions across 30 channels of EEG signals corrected by artifact removal methods during left-hand motor imagery. (B) Cortical surface topographic maps of $\mu$ band power obtained from ART-corrected EEG data, where the left- and right panels are maps in superior and inferior views, respectively. (C) Comparative classification accuracies of a two-class motor-imagery-based BCI using various artifact removal methods. These findings are derived from 2,480 and 2,438 EEG samples for left- and right-hand motor imagery tasks, respectively, across 109 recording sessions.

Figure 5A depicts the $\mu$ band (8–12 Hz) power distribution over a 30-channel EEG during right-hand motor imagery, presented in topographic maps generated using a 256-point fast Fourier transform for time–frequency transformation. Without artifact removal, the EEG displays expected power decreases in the $\mu$ band around the motor area over the contralateral side (left hemisphere), aligning with findings from prior studies [63]. The results processed with IC-U-Net and ART demonstrate their effectiveness not only in correcting potential artifacts but also in retaining key brain signatures indicative of neural activity during right-hand motor imagery. However, the three single-channel methods, ResCNN, DuoCL, and GCTNet, fail to retain such spatial patterns. Applying the source localization technique ROIconnect [61] to ART-corrected data from 2,480 right-hand motor imagery samples reveals a decrease in $\mu$ band power within the contralateral hemisphere. Notably, this reduction is observed in the superior frontal, paracentral, and superior parietal regions, as illustrated in Figure 5B. This result indicates that the ART model effectively preserves the integrity of EEG data without introducing distortions.

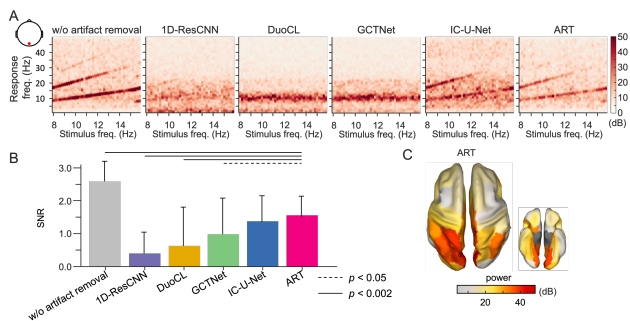

**Fig. 6.** Efficacy of artifact removal techniques on SSVEP responses. (A) Spectral powers of SSVEP at the Oz channel, corrected using various techniques. (B) SNR for SSVEP responses at Oz, averaged across 40 stimulus frequencies, with statistical analysis focusing on differences between ART and other methods. (C) Cortical surface topographic maps revealing EEG power distributions from SSVEP data processed by the ART method, presented in superior (left panel) and inferior (right panel) views. Note that these results are averaged across over 70 EEG recording sessions.

Next, this study investigates whether the performance of a motor imagery-based BCI system improves with data processed for artifact removal, specifically in classifying left-hand versus right-hand motor imagery. The BCI system follows the approach outlined in [64], using CSP for feature extraction and dimensionality reduction, retaining four CSP variance features, followed by linear discriminant analysis for classification. The average classification accuracy is computed from 10 holdout validation runs, each with an 80% training and 20% test sample split. As shown in Figure 5C, the accuracy for the BCI system without artifact removal processing is 0.55±0.05. With the incorporation of IC-U-Net and ART models, the BCI system's performance significantly increases to 0.69±0.04 (IC-U-Net) and 0.76±0.02 (ART), both with $p$-values < 0.001. However, 1D-ResCNN (0.56±0.03), DuoCL (0.54±0.02), and GCTNet (0.54±0.02) do not exhibit any significant improvement in BCI performance.

*SSVEP task*

The second selected popular BCI paradigm is the SSVEP task [50]. The dataset utilized (Table 1) contains stimulus frequencies ranging from 8.0 Hz to 15.8 Hz, encompassing 40 frequencies in intervals of 0.2 Hz. During SSVEP experiments, participants usually maintain a stable position, which helps minimize the introduction of artifacts. This stability often leads to a satisfactory SNR, reducing the need for extensive preprocessing, including artifact removal. Therefore, this analysis is aimed at investigating the effectiveness of each artifact removal technique in maintaining the original signal characteristics associated with SSVEP.

Figure 6A illustrates the spectral power of SSVEP derived from the Oz channel, associated with visual processing, after being corrected by various artifact removal techniques. Figure 6B shows the corresponding SNR values, comparing the results before and after applying various artifact removal techniques. In the absence of artifact removal, the EEG signals exhibit oscillatory activity with increased power at the frequency of the stimulus and its harmonics. The overall SNR for this condition is 2.587±0.623. Following artifact removal, the data processed by ART and IC-U-Net techniques yield SNRs of 1.551±0.585 and 1.373±0.817, respectively. Although these techniques result in some loss of signal clarity, they still retain the critical information present in the SSVEP harmonics. Figure 6C illustrates that SSVEP patterns, triggered by visual stimuli, remain detectable at the source level even after processing with the ART method. This preservation of EEG activity, synchronized with the stimulus frequency, manifests as power increases predominantly in the cuneus and superiorparietal lobule, highlighting the effectiveness of the ART in maintaining crucial signal characteristics. Conversely, 1D-ResCNN, DuoCL, and GCTNet show a marked inability to preserve SSVEP responses, with SNRs of 0.393±0.905, 0.622±1.767, and 0.978±1.481, respectively.

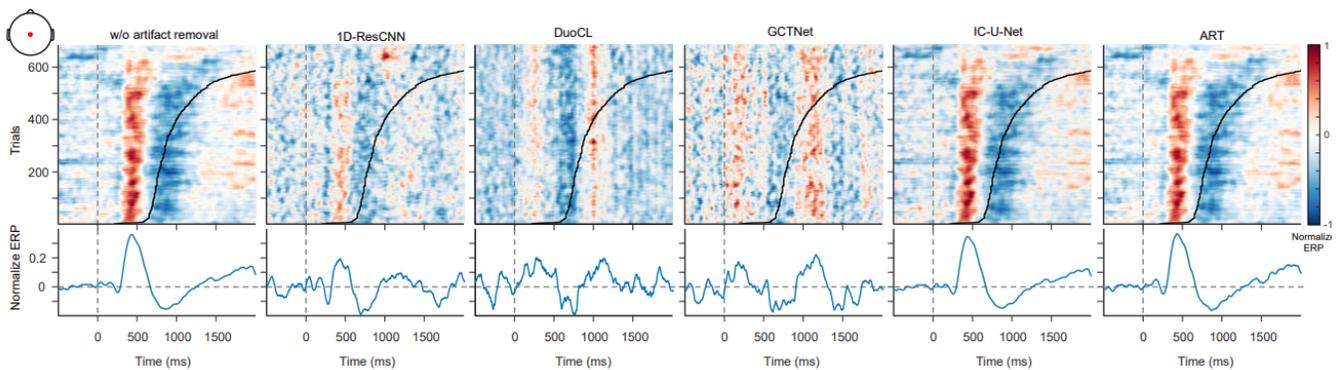

**Fig. 7.** Event-related analysis on EEG signals with artifact removal techniques. Each EEG trial is segmented to be time-locked to the onset of the departure event, capturing 500 ms before and 2,000 ms after the event. The upper panels illustrate ERP images sorted by reaction time to a lane-departure event, arranged from fast to slow responses, with the dashed line indicating the onset of the departure event and solid line marking the participant's response. The lower panels present the average ERP patterns across trials. The dashed and solid lines indicate the onsets of departure event and participant's responses, respectively. These results are derived from the Cz channel in a sustained-attention driving task.

*Sustained-attention driving task*

The third selected EEG dataset used to validate the effectiveness of the artifact removal techniques corresponds to the sustained-attention driving task [50]. The dataset pertains to an event-related lane-departure paradigm in a simulated vehicle mounted on a six-degree-of-freedom motion platform. The goal of this validation is to conduct a classical event-related potential (ERP) analysis to determine if stereotypical brain activity related to attention and decision-making, specifically the P300 component [65], remains present in the EEG signals after artifact removal. The EEG signals are segmented to be time-locked to critical events, allowing for the examination of changes in neural activity in response to these events.

The leftmost panel of Figure 7A illustrates a clear positive deflection approximately 437 ms after the lane-departure event, with an SNR of 17.220±0.268. This phenomenon is also observed in signals processed by IC-U-Net and ART, with comparable SNRs of 17.182±0.267 and 17.209±0.267, respectively. However, post-processed signals by 1D-ResCNN, DuoCL, and GCTNet show declined SNRs of 13.677±0.461, 14.440±0.428, and 12.208±0.660, respectively.

*Non-brain component suppression*

The final evaluation involves assessing the capability of these artifact removal techniques to suppress various artifacts using 10 EEG datasets [49, 50, 52, 54-56, 59, 60, 66], as listed in Table 1, excluding the first two datasets that were used to train the models. This comprehensive analysis includes decomposing and examining the brain and non-brain components of processed EEG signals through ICA and ICLabel. Tables S2 and S3 provide detailed numbers and likelihoods of decomposable brain components across various datasets. The range of identifiable brain components extends from 2.2±1.4 in the simultaneous EEG-fMRI dataset [54] to 17.0±8.6 in the motor imagery dataset from BCI Competition IV [66]. As anticipated, artifacts predominantly from eye and muscle activities are prominent, with up to 3.7±1.5 components in the simulated driving dataset [51] and 10.9±5.9 components in the motor imagery (BCI2000) dataset [49]. Additionally, indeterminate noises or components with multiple sources, categorized as "Other" [46], are significant in datasets like simultaneous EEG-fMRI [54] and speech-in-noise [56], with 21.1±2.7 and 22.2±3.7 components, respectively.

TABLE 3
NUMBERS OF NON-BRAIN COMPONENTS

| Dataset | # Channels[†] | w/o | 1D-ResCNN | DuoCL | GCTNet | IC-U-Net | IC-U-Net++ | IC-U-Net-Attn | ART |
|---|---|---|---|---|---|---|---|---|---|
| Simulated Driving [51] | 30 | 17.1±3.3 | 7.3±2.9 | 1.7±2.4 | 3.2±1.7 | 3.0±1.8 | 2.1±0.9 | 2.6±1.3 | 3.2±1.9 |
| Motor Imagery (BCI2000) [49] | 30 | 18.3±4.8 | 20.5±1.6 | 0.2±0.2 | 9.1±1.9 | 3.8±1.9 | 3.1±1.2 | 3.8±1.3 | 2.9±2.2 |
| Visual Speller (SSVEP) [50] | 30 | 12.7±2.3 | 17.2±2.2 | 0.2±0.5 | 7.5±1.9 | 1.7±0.8 | 2.6±1.2 | 3.1±1.1 | 1.8±0.9 |
| Simultaneous EEG-fMRI [54] | 28 | 25.8±2.6 | 22.2±2.3 | 9.6±2.4 | 17.6±2.9 | 12.8±2.0 | 13.5±2.5 | 6.8±2.1 | 17.8±3.0 |
| Eyes-Closed Resting-State [60] | 19 | 5.9±1.3 | 2.8±1.3 | 1.1±0.5 | 1.6±0.9 | 3.9±1.1 | 0.4±0.6 | 1.6±0.8 | 3.5±1.6 |
| Error Processing [55] | 30 | 13.0±1.7 | 5.9±1.0 | 1.2±0.5 | 2.1±1.0 | 4.1±1.3 | 2.2±1.8 | 4.3±2.2 | 1.5±0.8 |
| Speech-in-Noise [56] | 30 | 25.1±3.4 | 4.9±1.5 | 4.5±1.9 | 1.6±0.9 | 4.3±2.2 | 3.1±1.3 | 5.1±1.6 | 2.9±1.8 |
| Motor Imagery (BCI Competition IV) [66] | 30 | 13.0±7.8 | 8.7±5.4 | 1.9±1.3 | 1.6±1.0 | 6.9±4.5 | 3.0±2.2 | 3.3±1.7 | 7.1±5.8 |
| Go/No-Go [59] | 21 | 11.6±2.2 | 4.7±1.2 | 1.8±0.4 | 2.4±0.9 | 1.8±0.9 | 1.3±0.9 | 1.0±0.5 | 1.1±0.7 |
| Oddball [52] | 30 | 15.5±4.6 | 4.5±2.0 | 1.4±0.4 | 1.4±0.7 | 3.3±2.3 | 2.7±2.7 | 5.8±2.3 | 6.9±4.1 |

[†]For datasets containing more than 30 channels, electrodes were mapped to match those specified in the training data's configuration as closely as possible.

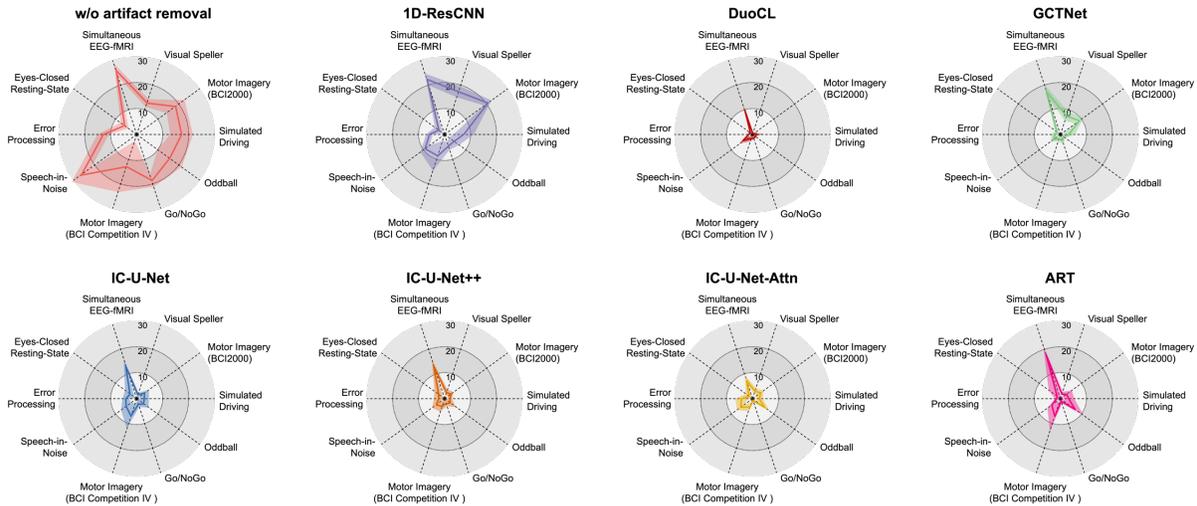

**Fig. 8.** Non-brain component counts in EEG data processed by different artifact removal techniques. This spider plot displays axes representing 11 selected datasets, as listed in Table 1. Each axis has a maximum scale of 30 components, illustrating the efficacy of each technique in reducing non-brain components across the datasets.

Table 3 summarizes the number of detectable non-brain components in EEG signals processed by various artifact removal techniques. Figure 8 complements this with spider charts summarizing the number of detectable non-brain components before and after processing with each technique across multiple datasets. Smaller polygon areas in these charts indicate better artifact suppression. Results demonstrate significant artifact reduction, with the polygon area shrinking from 682.1 (without processing) to 341.5 (1D-ResCNN), 10.0 (DuoCL), 82.5 (GCTNet), 50.3 (IC-U-Net), 35.6 (IC-U-Net++), 33.0 (IC-U-Net-Attn), and 37.2 (ART), calculated using the shoelace formula [67]. DuoCL emerges as the most effective, closely followed by IC-U-Net++, IC-U-Net-Attn, and ART. Although DuoCL demonstrates high efficiency, it might distort essential brain signatures, as indicated in assessments based on MSE, SSVEP, motor imagery, BCI, and ERP.

*Computational efficiency comparison*

Table 4 compares seven algorithms based on model complexity and computational efficiency under standardized conditions—processing 4-s EEG segments across 30 channels at a sampling rate of 256 Hz. All models complete processing within one second, suitable for near real-time EEG analysis. IC-U-Net and IC-U-Net++ feature low parameter counts and very short inference times, underscoring the advantages of pure CNN architectures in speed and efficiency. ART, despite slightly higher parameter numbers than IC-U-Net, also maintains quick inference times, albeit slightly longer. In contrast, 1D-ResCNN, DuoCL, and GCTNet exhibit longer inference times due to their channel-by-channel signal processing approach.

## V. Discussion

The proposed EEG denoising model, ART, demonstrates exceptional performance, surpassing IC-U-Net [38, 47], 1D-ResCNN[21], DuoCL [31], and GCTNet [34] across several metrics. ART achieves lower MSE, indicating its capability not only to generate but also to accurately recover signals. Its high SNR and preservation of ERP waveforms and accurate source localization further underscore its ability to remove artifacts while retaining essential brain activities. Additionally, when signals corrected by ART are used to train BCI, the classification accuracy is significantly improved, showcasing its practical effectiveness in enhancing BCI performance. The following sections analyze the behaviors and limitations of the ART model to elucidate areas where the model excels and identify potential challenges or shortcomings that could be addressed in future iterations or applications.

### A. Model Behaviors

Multi-head attention mechanisms have demonstrated significant utility [68-70] in various fields beyond EEG

TABLE 4
MODEL COMPLEXITY AND INFERENCE TIME

| Method | Parameter (million) | Inference time (s) |
|---|---|---|
| 1D-ResCNN | 8.46 | 0.46 |
| DuoCL | 36.81 | 0.93 |
| GCTNet | 29.38 | 0.69 |
| IC-U-Net | 2.67 | **0.02** |
| IC-U-Net++ | **1.75** | **0.02** |
| IC-U-Net-Attn | 46.37 | 0.93 |
| ART | 2.51 | 0.19 |

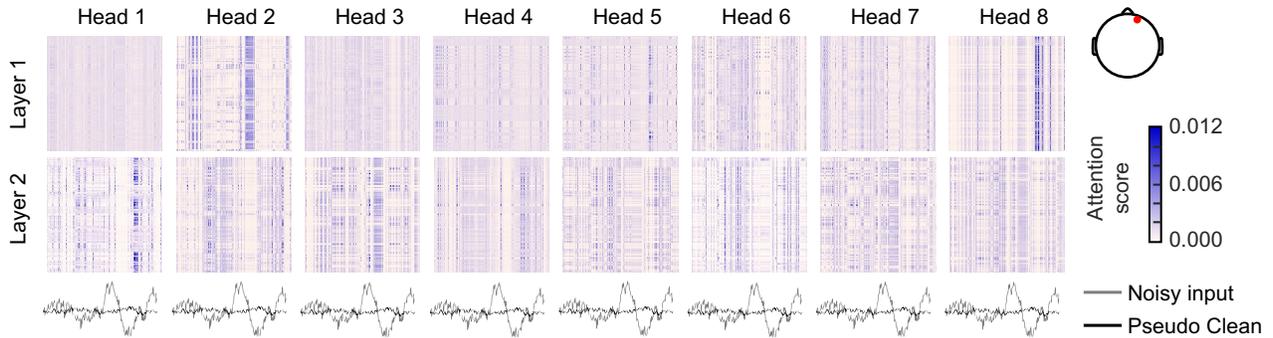

**Fig. 9.** Heatmaps of attention scores from multiple heads in the first two layers, analyzing a signal segment from the Fp2 channel, known for contamination by eye artifacts.

studies. In sentiment analysis, they help models focus on salient features within the text that can contribute to the overall sentiment of a sentence or document [71]. Similarly, in time-series forecasting, multi-head attention enhances the ability of models to focus on crucial temporal features to effectively predict future values [72]. Applying these insights into the ART model, an in-depth exploration of its multi-head attention mechanisms becomes crucial to clarify how the model processes and reconstructs artifact-free EEG signals, potentially illuminating the internal dynamics of feature extraction. The attention maps from multiple heads in the initial two layers are depicted in Figure 9. Each map represents a square matrix of attention scores, derived from the dot products of the encoder keys and decoder queries. The varied patterns across these maps suggest that different heads focus on distinct representational subspaces and specific positions within the EEG data. Notably, larger attention scores are often observed in segments heavily affected by artifacts, such as in the second head of the first layer, indicating the model's strategic choices in data retention and reconstruction.

The choice of target sequence critically influences the performance of the ART in artifact removal. This study examines three distinct target sequence configurations within the ART model to determine their influence on EEG artifact removal effectiveness. The third configuration of the ART model, using noisy input as the target sequence, remarkably outperforms the configurations with pseudo-clean and zero matrix inputs. This outcome is aligned with insights from deep-learning research suggesting that incorporating moderate noise during training can enhance the robustness and generalizability of neural networks [73, 74]. Injecting noise into training data has been shown to improve feature extraction capabilities by forcing the model to learn to distinguish relevant patterns from irrelevant noise [75]. Additionally, noise injection may act as a form of regularization, helping prevent overfitting to overly simplistic or noise-free training data [76]. This approach mimics more realistic conditions that a model might encounter in real-world applications, thereby preparing it to perform more effectively in noisy environments. These findings underscore the potential of noise training to create neural networks that more accurately mirror neural responses to complex, noisy stimuli, suggesting promising directions for future research in EEG artifact removal and other applications involving high levels of data variability.

*B. Limitations and Suggestions*

While the ART model demonstrates superiority in reconstructing EEG signals and mitigating various artifacts, several limitations must be addressed in future work. First, ART is among the few EEG denoising techniques capable of removing artifacts from multichannel EEG time-series in a single step. However, to achieve optimal results, the number of channels and their locations must closely mirror the initial settings used in the ART model. We recommend employing a channel mapping strategy before applying ART. If channel locations differ from those in the ART training dataset, as in setups with more than 30 channels or different EEG montages, channel mapping can help align channels with those in the training dataset based on proximity. In cases with fewer than 30 channels, spherical spline interpolation can be applied to estimate missing data using neighboring electrode information. Second, while ART effectively suppresses artifacts from eyes, heart, and muscles, as well as line and channel noise, it does not guarantee the removal of all noise types, such as gradient artifacts from EEG-fMRI studies, which remain a challenge. In this study, decomposed component analysis often classifies unaddressed signals as "Other," indicating the presence of indeterminate noise. For such specialized artifacts, preprocessing with targeted tools before applying ART is advisable. The effects of repeated application of artifact removal methods, including ART, on EEG signals, have not been extensively studied. Investigating the impact of multiple corrections on signal integrity is essential. Moreover, ART's compatibility with other artifact removal techniques remains to be assessed. Further studies are needed to explore the efficacy and validity of integrating ART with existing methods.

## V. CONCLUSIONS

This study demonstrates the efficacy of the ART model, a novel EEG denoising technique that leverages transformer architecture to adeptly handle multichannel EEG data. Integrating self-attention into the previously proposed IC-U-

Net model, with the inclusion of positional encoding, marks a pivotal first step in using both temporal and spatial information within the EEG time-series. This foundational enhancement is instrumental for the subsequent development and success of the ART model. ART's performance is superior to the competing deep-learning-based methods, evidenced by its substantial reductions in MSE, improvements in SNR, and enhanced BCI application performance. Moreover, ART's ability to preserve essential brain activity during artifact removal and its adaptability to various EEG configurations underscore its potential as a formidable tool in neuroscientific research. The insights gained from this study will guide further innovations in EEG processing, aiming for more accurate and reliable artifact removal methods.

## REFERENCES

[1] R. Abiri, S. Borhani, E. W. Sellers, Y. Jiang, and X. Zhao, "A comprehensive review of EEG-based brain–computer interface paradigms," *Journal of Neural Engineering,* vol. 16, no. 011001, 2019.

[2] E. Jungnickel and K. Gramann, "Mobile brain/body imaging (MoBI) of physical interaction with dynamically moving objects," *Frontiers in Human Neuroscience,* vol. 10, no. 306, pp. 1-15, 2016.

[3] X. Jiang, G. B. Bian, and Z. Tian, "Removal of artifacts from EEG signals: A review," *Sensors,* vol. 19, no. 5, 2019.

[4] W. Mumtaz, S. Rasheed, and A. Irfan, "Review of challenges associated with the EEG artifact removal methods," *Biomedical Signal Processing and Control,* vol. 68, no. 102741, 2021.

[5] J. Minguillon, M. A. Lopez-Gordo, and F. Pelayo, "Trends in EEG-BCI for daily-life: Requirements for artifact removal," *Biomedical Signal Processing and Control,* vol. 31, pp. 407-418, 2017.

[6] S. Puthusserypady and T. Ratnarajah, "Robust adaptive techniques for minimization of EOG artefacts from EEG signals," *Signal Processing,* vol. 86, no. 9, pp. 2351-2363, 2006.

[7] S. D. Muthukumaraswamy, "High-frequency brain activity and muscle artifacts in MEG/EEG: A review and recommendations," *Frontiers in Human Neuroscience,* vol. 7, 2013.

[8] C. Dai, J. Wang, J. Xie, W. Li, Y. Gong, and Y. Li, "Removal of ECG artifacts from EEG using an effective recursive least square notch filter," *IEEE Access,* vol. 7, pp. 158872-158880, 2019.

[9] E. H. T. Shad, M. Molinas, and T. Ytterdal, "Impedance and noise of passive and active dry EEG electrodes: A review," *IEEE Sensors Journal,* vol. 20, no. 24, pp. 14565-14577, 2020.

[10] Y. Roy, H. Banville, I. Albuquerque, A. Gramfort, T. H. Falk, and J. Faubert, "Deep learning-based electroencephalography analysis: A systematic review," *Journal of Neural Engineering,* vol. 16, no. 051001, 2019.

[11] V. J. Lawhern, A. J. Solon, N. R. Waytowich, S. M. Gordon, C. P. Hung, and B. J. Lance, "EEGNet: A compact convolutional neural network for EEG-based brain–computer interfaces," *Journal of Neural Engineering,* vol. 15, no. 056013, 2018.

[12] H. Altaheri, G. Muhammad, M. Alsulaiman, S. U. Amin, G. A. Altuwaijri, W. Abdul *et al.*, "Deep learning techniques for classification of electroencephalogram (EEG) motor imagery (MI) signals: A review," *Neural Computing and Applications,* vol. 35, no. 20, pp. 14681-14722, 2023.

[13] J. Li, Z. Struzik, L. Zhang, and A. Cichocki, "Feature learning from incomplete EEG with denoising autoencoder," *Neurocomputing,* vol. 165, pp. 23-31, 2015.

[14] S. Stober, A. Sternin, A. M. Owen, and J. A. Grahn, "Deep feature learning for EEG recordings," *arXiv:1511.04306,* 2015.

[15] J. T. Schwabedal, J. C. Snyder, A. Cakmak, S. Nemati, and G. D. Clifford, "Addressing class imbalance in classification problems of noisy signals by using Fourier transform surrogates," *arXiv:1806.08675,* 2018.

[16] F. Wang, S.-h. Zhong, J. Peng, J. Jiang, and Y. Liu, "Data augmentation for EEG-based emotion recognition with deep convolutional neural networks," in *MultiMedia Modeling: 24th International Conference, MMM 2018, Bangkok, Thailand, February 5-7, 2018, Proceedings, Part II 24*, 2018: Springer, pp. 82-93.

[17] B. Yang, K. Duan, and T. Zhang, "Removal of EOG artifacts from EEG using a cascade of sparse autoencoder and recursive least squares adaptive filter," *Neurocomputing,* vol. 214, pp. 1053-1060, 2016.

[18] B. Yang, K. Duan, C. Fan, C. Hu, and J. Wang, "Automatic ocular artifacts removal in EEG using deep learning," *Biomedical Signal Processing and Control,* vol. 43, pp. 148-158, 2018.

[19] K. He, X. Zhang, S. Ren, and J. Sun, "Deep residual learning for image recognition," in *Proceedings of the IEEE Conference on Computer Vision and Pattern Recognition*, 2016, pp. 770-778.

[20] Y. LeCun, B. Boser, J. S. Denker, D. Henderson, R. E. Howard, W. Hubbard, and L. D. Jackel, "Backpropagation applied to handwritten zip code recognition," *Neural Computation,* vol. 1, no. 4, pp. 541-551, 1989.

[21] W. Sun, Y. Su, X. Wu, and X. Wu, "A novel end-to-end 1D-ResCNN model to remove artifact from EEG signals," *Neurocomputing,* vol. 404, pp. 108-121, 2020.

[22] H. Zhang, C. Wei, M. Zhao, Q. Liu, and H. Wu, "A novel convolutional neural network model to remove muscle artifacts from EEG," in *ICASSP 2021-2021 IEEE International Conference on Acoustics, Speech and Signal Processing (ICASSP)*, 2021: IEEE, pp. 1265-1269.

[23] I. Goodfellow, J. Pouget-Abadie, M. Mirza, B. Xu, D. Warde-Farley, S. Ozair *et al.*, "Generative adversarial nets," *Advances in Neural Information Processing Systems,* vol. 27, 2014.

[24] K. G. Hartmann, R. T. Schirrmeister, and T. Ball, "EEG-GAN: Generative adversarial networks for electroencephalograhic (EEG) brain signals," *arXiv:1806.01875,* 2018.

[25] T.-j. Luo, Y. Fan, L. Chen, G. Guo, and C. Zhou, "EEG signal reconstruction using a generative adversarial network with wasserstein distance and temporal-spatial-frequency loss," *Frontiers in Neuroinformatics,* vol. 14, 2020.

[26] P. Sawangjai, M. Trakulruangroj, C. Boonnag, M. Piriyajitakonkij, R. K. Tripathy, T. Sudhawiyangkul, and T. Wilaiprasitporn, "EEGANet: Removal of ocular artifacts from the EEG signal using generative adversarial networks," *IEEE Journal of Biomedical and Health Informatics,* vol. 26, no. 10, pp. 4913-4924, 2021.

[27] A. Sherstinsky, "Fundamentals of recurrent neural network (RNN) and long short-term memory (LSTM) network," *Physica D: Nonlinear Phenomena,* vol. 404, 2020.

[28] S. Hochreiter and J. Schmidhuber, "Long short-term memory," *Neural Computation,* vol. 9, no. 8, pp. 1735-1780, 1997.

[29] H. Zhang, M. Zhao, C. Wei, D. Mantini, Z. Li, and Q. Liu, "EEGdenoiseNet: A benchmark dataset for deep learning solutions of EEG denoising," *Journal of Neural Engineering,* vol. 18, no. 5, 2021.

[30] J. R. McIntosh, J. Yao, L. Hong, J. Faller, and P. Sajda, "Ballistocardiogram artifact reduction in simultaneous EEG-fMRI using deep learning," *IEEE Transactions on Biomedical Engineering,* vol. 68, no. 1, pp. 78-89, 2021.

[31] T. Gao, D. Chen, Y. Tang, Z. Ming, and X. Li, "EEG reconstruction with a dual-scale CNN-LSTM model for deep artifact removal," *IEEE Journal of Biomedical and Health Informatics,* vol. 27, no. 3, pp. 1283-1294, 2022.

[32] A. Vaswani, N. Shazeer, N. Parmar, J. Uszkoreit, L. Jones, A. N. Gomez *et al.*, "Attention is all you need," *Advances in Neural Information Processing Systems,* vol. 30, 2017.

[33] X. Pu, P. Yi, K. Chen, Z. Ma, D. Zhao, and Y. Ren, "EEGDnet: Fusing non-local and local self-similarity for EEG signal denoising with transformer," *Computers in Biology and Medicine,* vol. 151, Part A, no. 106248, 2022.

[34] J. Yin, A. Liu, C. Li, R. Qian, and X. Chen, "A GAN guided parallel CNN and transformer network for EEG denoising," *IEEE Journal of Biomedical and Health Informatics,* vol. (Early Access), pp. 1-12, 2023.

[35] J. Cao, Y. Zhao, X. Shan, H.-l. Wei, Y. Guo, L. Chen *et al.*, "Brain functional and effective connectivity based on electroencephalography recordings: A review," *Human Brain Mapping,* vol. 43, no. 2, pp. 860-879, 2022.

[36] X. Liu, Y. Shen, J. Liu, J. Yang, P. Xiong, and F. Lin, "Parallel spatial–temporal self-attention CNN-based motor imagery classification for BCI," *Frontiers in Neuroscience,* vol. 14, 2020.

[37] F. Lotte, L. Bougrain, A. Cichocki, M. Clerc, M. Congedo, A. Rakotomamonjy, and F. Yger, "A review of classification algorithms for EEG-based brain–computer interfaces: A 10 year update," *Journal of Neural Engineering,* vol. 15, no. 031005, 2018.

[38] C.-H. Chuang, K.-Y. Chang, C.-S. Huang, and T.-P. Jung, "IC-U-Net: A U-Net-based denoising autoencoder using mixtures of independent

components for automatic EEG artifact removal," *NeuroImage,* vol. 263, p. 119586, 2022.
[39] O. Ronneberger, P. Fischer, and T. Brox, "U-net: Convolutional networks for biomedical image segmentation," in *Medical Image Computing and Computer-assisted Intervention–MICCAI 2015*, 2015, Munich, Germany: Springer, pp. 234-241.
[40] L. Feng, C. Cheng, M. Zhao, H. Deng, and Y. Zhang, "EEG-based emotion recognition using spatial-temporal graph convolutional LSTM With attention mechanism," *IEEE Journal of Biomedical and Health Informatics,* vol. 26, no. 11, pp. 5406-5417, 2022.
[41] Y. Du, Y. Xu, X. Wang, L. Liu, and P. Ma, "EEG temporal–spatial transformer for person identification," *Scientific Reports,* vol. 12, no. 14378, 2022.
[42] A. Dosovitskiy, L. Beyer, A. Kolesnikov, D. Weissenborn, X. Zhai, T. Unterthiner *et al.*, "An image is worth 16x16 words: Transformers for image recognition at scale," *arXiv:2010.11929*, 2020.
[43] A. Radford, J. W. Kim, T. Xu, G. Brockman, C. McLeavey, and I. Sutskever, "Robust speech recognition via large-scale weak supervision," in *International Conference on Machine Learning*, 2023: PMLR, pp. 28492-28518.
[44] S. Makeig, A. Bell, T.-P. Jung, and T. Sejnowski, "Independent component analysis of electroencephalographic data," *Advances in Neural Information Processing Systems,* vol. 8, no. 8, pp. 145-151, 1996.
[45] T. P. Jung, S. Makeig, C. Humphries, T. W. Lee, M. J. McKeown, V. Iragui, and T. J. Sejnowski, "Removing electroencephalographic artifacts by blind source separation," *Psychophysiology,* vol. 37, no. 2, pp. 163-78, 2000.
[46] L. Pion-Tonachini, K. Kreutz-Delgado, and S. Makeig, "ICLabel: An automated electroencephalographic independent component classifier, dataset, and website," *NeuroImage,* vol. 198, pp. 181-197, 2019.
[47] K. Y. Chang, Y. C. Huang, and C. H. Chuang, "Enhancing EEG artifact removal efficiency by introducing dense skip connections to IC-U-Net," in *2023 45th Annual International Conference of the IEEE Engineering in Medicine & Biology Society (EMBC)*, 24-27 July 2023 2023, pp. 1-4.
[48] Z. Zhou, M. M. Rahman Siddiquee, N. Tajbakhsh, and J. Liang, "Unet++: A nested u-net architecture for medical image segmentation," in *Deep Learning in Medical Image Analysis, DLMIA 2018*, Granada, Spain, 2018: Springer, pp. 3-11.
[49] G. Schalk, D. J. McFarland, T. Hinterberger, N. Birbaumer, and J. R. Wolpaw, "BCI2000: A general-purpose brain-computer interface (BCI) system," *IEEE Transactions on Biomedical Engineering,* vol. 51, no. 6, pp. 1034-43, 2004.
[50] B. Liu, X. Huang, Y. Wang, X. Chen, and X. Gao, "BETA: A large benchmark database toward SSVEP-BCI application," *Frontiers in Neuroscience,* vol. 14, 2020.
[51] Z. Cao, C.-H. Chuang, J.-K. King, and C.-T. Lin, "Multi-channel EEG recordings during a sustained-attention driving task," *Scientific Data,* vol. 6, no. 19, pp. 1-8, 2019.
[52] Y. Li, B. Yang, Z. Wang, R. Huang, X. Lu, X. Bi, and S. Zhou, "EEG assessment of brain dysfunction for patients with chronic primary pain and depression under auditory oddball task," *Frontiers in Neuroscience,* vol. 17, 2023.
[53] Y. J. Kim, M. Grabowecky, K. A. Paller, and S. Suzuki, "Differential roles of frequency-following and frequency-doubling visual responses revealed by evoked neural harmonics," *Journal of Cognitive Neuroscience,* vol. 23, no. 8, pp. 1875-86, 2011.
[54] Y. Gu, L. E. Sainburg, F. Han, and X. Liu, "Simultaneous EEG and functional MRI data during rest and sleep from humans," *Data in Brief,* vol. 48, no. 109059, 2023.
[55] R. Chavarriaga and J. d. R. Millan, "Learning from EEG error-related potentials in noninvasive brain-computer interfaces," *IEEE Transactions on Neural Systems and Rehabilitation Engineering,* vol. 18, no. 4, pp. 381-388, 2010.
[56] M. P. Broderick, A. J. Anderson, G. M. Di Liberto, M. J. Crosse, and E. C. Lalor, "Electrophysiological correlates of semantic dissimilarity reflect the comprehension of natural, narrative speech," *Current Biology,* vol. 28, no. 5, pp. 803-809.e3, 2018.
[57] R. J. Kobler, A. I. Sburlea, C. Lopes-Dias, A. Schwarz, M. Hirata, and G. R. Müller-Putz, "Corneo-retinal-dipole and eyelid-related eye artifacts can be corrected offline and online in electroencephalographic and magnetoencephalographic signals," *NeuroImage,* vol. 218, no. 117000, 2020.
[58] P. Sajda, A. Gerson, K. R. Muller, B. Blankertz, and L. Parra, "A data analysis competition to evaluate machine learning algorithms for use in brain-computer interfaces," *IEEE Transactions on Neural Systems and Rehabilitation Engineering,* vol. 11, no. 2, pp. 184-185, 2003.
[59] A. Delorme, S. Makeig, M. Fabre-Thorpe, and T. Sejnowski, "From single-trial EEG to brain area dynamics," *Neurocomputing,* vol. 44-46, pp. 1057-1064, 2002.
[60] A. Miltiadous, K. D. Tzimourta, T. Afrantou, P. Ioannidis, N. Grigoriadis, D. G. Tsalikakis *et al.*, "A dataset of scalp EEG recordings of Alzheimer's disease, frontotemporal dementia and healthy subjects from routine EEG," *Data,* vol. 8, no. 6, 2023.
[61] F. Pellegrini, A. Delorme, V. Nikulin, and S. Haufe, "Identifying good practices for detecting inter-regional linear functional connectivity from EEG," *NeuroImage,* vol. 277, no. 120218, 2023.
[62] F. Lotte, "A Tutorial on EEG Signal-processing Techniques for Mental-state Recognition in Brain–Computer Interfaces," in *Guide to Brain-Computer Music Interfacing*, E. R. Miranda and J. Castet Eds. London: Springer London, 2014, pp. 133-161.
[63] G. Pfurtscheller and F. H. Lopes da Silva, "Event-related EEG/MEG synchronization and desynchronization: basic principles," *Clinical Neurophysiology,* vol. 110, no. 11, pp. 1842-57, 1999.
[64] M. Billinger. "Motor imagery decoding from EEG data using the common spatial pattern (CSP)." n.d. https://mne.tools/stable/auto_examples/decoding/decoding_csp_eeg.html (accessed 25 July, 2023).
[65] T. W. Picton, "The P300 wave of the human event-related potential," *Journal of Clinical Neurophysiology,* vol. 9, no. 4, pp. 456-79, 1992.
[66] B. Blankertz, G. Dornhege, M. Krauledat, K.-R. Müller, and G. Curio, "The non-invasive Berlin brain-computer interface: Fast acquisition of effective performance in untrained subjects," *NeuroImage,* vol. 37, no. 2, pp. 539-550, 2007.
[67] E. W. Weisstein. "Shoelace Formula." n.d. https://mathworld.wolfram.com/ShoelaceFormula.html (accessed 21 Jun, 2024).
[68] H. Gong, Y. Tang, J. M. Pino, and X. Li, "Pay better attention to attention: head selection in multilingual and multi-domain sequence modeling," presented at the Proceedings of the 35th International Conference on Neural Information Processing Systems, 2024.
[69] E. Voita, D. Talbot, F. Moiseev, R. Sennrich, and I. Titov, "Analyzing multi-head self-attention: Specialized heads do the heavy lifting, the rest can be pruned," *arXiv:1905.09418*, 2019.
[70] J.-B. Cordonnier, A. Loukas, and M. Jaggi, "Multi-head attention: Collaborate instead of concatenate," *arXiv:2006.16362*, 2020.
[71] F. Long, K. Zhou, and W. Ou, "Sentiment analysis of text based on bidirectional LSTM with multi-head attention," *IEEE Access,* vol. 7, pp. 141960-141969, 2019.
[72] H. Abbasimehr and R. Paki, "Improving time series forecasting using LSTM and attention models," *Journal of Ambient Intelligence and Humanized Computing,* vol. 13, no. 1, pp. 673-691, 2022.
[73] S. L. Smith, E. Elsen, and S. De, "On the generalization benefit of noise in stochastic gradient descent," presented at the Proceedings of the 37th International Conference on Machine Learning, 2020.
[74] S. Zheng, Y. Song, T. Leung, and I. Goodfellow, "Improving the robustness of deep neural networks via stability training," in *Proceedings of the IEEE Conference on Computer Vision and Pattern Recognition*, 2016, pp. 4480-4488.
[75] S. Yin, C. Liu, Z. Zhang, Y. Lin, D. Wang, J. Tejedor *et al.*, "Noisy training for deep neural networks in speech recognition," *EURASIP Journal on Audio, Speech, and Music Processing,* vol. 2, 2015.
[76] H. Noh, T. You, J. Mun, and B. Han, "Regularizing deep neural networks by noise: its interpretation and optimization," presented at the Proceedings of the 31st International Conference on Neural Information Processing Systems, Long Beach, California, USA, 2017.

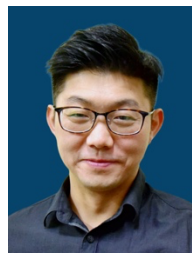

**Chun-Hsiang Chuang** received his Ph.D. in Electrical Engineering from National Chiao Tung University, Hsinchu, Taiwan, in 2014. He served as a Visiting Scholar at the Swartz Center for Computational Neuroscience, University of California,

San Diego, from 2012 to 2013. He then held a Lecturer position at the School of Software, University of Technology Sydney, Australia, from 2016 to 2017. He was appointed as an Assistant Professor in the Department of Computer Science and Engineering at National Taiwan Ocean University from 2018 to 2020. Between 2021 and 2023, he served as the Treasurer of the IEEE Computational Intelligence Society, Taipei Section. Currently, he is the Deputy Director and an Associate Professor at the Center for Education and Mind Sciences, College of Education, and the Institute of Information Systems and Applications, College of Electrical Engineering and Computer Science, National Tsing Hua University, Taiwan. Additionally, he is the Secretary of the Taipei Chapter of the IEEE Systems, Man, and Cybernetics Society. His research interests include artificial intelligence, brain-computer interfaces, virtual/augmented reality, neuroeducation, and neuroergonomics.

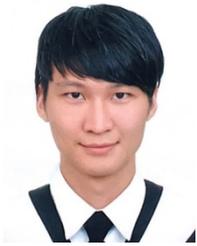

**Kong-Yi Chang** received his Master's degree in Computer Science from National Taiwan Ocean University, Keelung, Taiwan, in 2022. He is currently a Ph.D. candidate at the Institute of Information Systems and Applications, National Tsing Hua University. His research focuses on artificial intelligence, brain-computer interfaces, and signal processing.

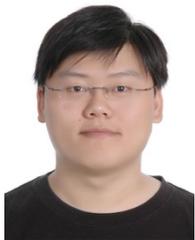

**Chih-Sheng Huang** received his Ph.D. from the Institute of Electrical Control Engineering at National Chiao Tung University, Hsinchu, Taiwan, in 2015. He is currently the Assistant Manager for Artificial Intelligence Research and Development in the Product Development Division at ELAN Microelectronics Corporation. Dr. Huang also holds joint appointments as an Assistant Professor at the College of Artificial Intelligence, National Yang Ming Chiao Tung University, and at the College of Electrical Engineering and Computer Science, National Taipei University of Technology. His research primarily focuses on the development of deep learning for edge computing and data mining for edge AI models.

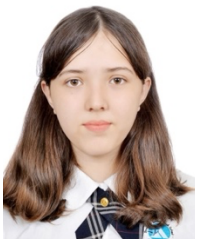

**Anne-Mei Bessas** is currently pursuing a Bachelor's degree at the Research Center for Education and Mind Sciences, National Tsing Hua University. Her research interests focus on EEG analysis and brain–computer interfaces.